\documentclass[prl,floats,aps,twocolumn,epsf,graphicx]{revtex4}
\usepackage{epsfig}
\usepackage{graphicx}

\begin{document}

\title{Lifetime of Unstable Hairy Black Holes}
\author{Shahar Hod}
\address{The Ruppin Academic Center, Emeq Hefer 40250, Israel}
\address{}
\address{The Hadassah Institute, Jerusalem 91010, Israel}
\date{\today}

\begin{abstract}

\ \ \ During the last two decades solutions of black holes with
various types of ``hair'' have been discovered. Remarkably, it has
been established that many of these hairy black holes are unstable--
under small perturbations the hair may collapse. While the static
sector of theories admitting hair is well explored by now, our
picture of the dynamical process of hair-shedding is still
incomplete. In this Letter we provide an important ingredient of the
nonlinear dynamics of hair collapse: we derive a universal lower
bound on the lifetime of hairy black holes. It is also shown that
the amount of hair outside of a black-hole horizon should be
fundamentally bounded.
\end{abstract}
\bigskip
\maketitle

Wheeler's dictum ``a black hole has no hair'' \cite{Whee} has played
a major role in the development of black-hole physics
\cite{BekTod,Nun}. The conjecture implies that black holes are
fundamental objects: they should be described by only a few
parameters, very much like atoms in quantum mechanics. The fact that
stationary black holes are specified by conserved charges which are
associated with a Gauss-like law (mass, charge, and angular
momentum) was proven explicitly in Einstein vacuum theory and
Einstein-Maxwell theory \cite{Wald}. Early no-hair theorems also
excluded scalar \cite{Chas}, massive vector \cite{BekVec}, and
spinor \cite{Hart} fields from the exterior of a stationary black
hole.

However, later day developments in particle physics have lead to the
somewhat surprising discovery of various types of ``hairy'' black
holes, the first of which were the ``colored black holes''
\cite{BizCol}. These are static black hole solutions of the
Einstein-Yang-Mills (EYM) theory that require for their complete
specification not only the value of the mass, but also an additional
integer (counting the number of nodes of the function characterizing
the Yang-Mills field outside the horizon) that is not, however,
associated with any conserved charge. Following the discovery of the
colored black holes, other hairy solutions have been found
\cite{Lavr,BizCham,Green}.

Remarkably, many of these hairy black-hole solutions were proven to
be unstable \cite{Stra,BiWa,Volkov}. This implies that under small
perturbations the black hole would lose its hair. Numerical studies
of hairy black-hole spacetimes \cite{Chop1,Chop2} have revealed two
distinct mechanisms of losing unstable hair: the hair may be
dispersed to infinity, leaving the original black hole virtually
unchanged, or it may collapse into the black hole, in which case the
original black hole gets bigger.

While the {\it static} sector of theories admitting hair is well
understood by now, we still luck a complete picture of the {\it
dynamical} processes by which perturbed, unstable hairy black holes
shed their hair. In particular, the fact that many of the hairy
black-hole solutions were found to be dynamically unstable promotes
the fundamental question: what is the lifetime of an unstable hairy
black hole?

In this Letter we derive a universal lower bound on the dynamical
lifetime, $\tau_{hair}$, of perturbed, unstable hairy black-hole
solutions. Our derivation is based on standard results of
information theory and universal thermodynamic considerations.

A fundamental question in quantum information theory is what is the
maximum rate at which information/entropy \cite{noteentinf} may be
changed by a signal of duration $\tau$ and energy $\Delta {\cal E}$.
The answer to this question is given by the Bekenstein-Bremermann
relation \cite{Bekinf,Bre,NoteCont}:

\begin{equation}\label{Eq1}
{{\Delta S} \over \tau} \leq \pi \Delta{\cal E} / \hbar \  .
\end{equation}
(We use gravitational units in which $G=c=k_B=1$.) We shall now use
this relation in the context of black-hole dynamics.

Accretion of perturbed, unstable hair into a black hole would
increase the black-hole mass \cite{wec} by an amount $\Delta{\cal
E}$ ($\Delta{\cal E} \leq {\cal M}_{hair}$, where ${\cal M}_{hair}$
is the mass of the black-hole hair outside the horizon). As a
consequence its area (entropy) would also increase. This implies
that information/entropy is actually flowing into the black hole
during this dynamical process.

One may now derive an upper bound on the dynamical rate at which the
hair collapse (or equivalently, an upper bound on the rate at which
the black-hole entropy increases due to hair ``swallowing''). For
simplicity we focus here on spherically symmetric spacetimes. The
mass $m(r_H)$ contained within the black-hole horizon is related to
the horizon's radius by $m(r_H)={1 \over 2}r_H$ \cite{Nun}, and the
mass of the outside hair is given by ${\cal M}_{hair}={\cal
M}_{total}-{1 \over 2}r_H$. It is worth mentioning that \cite{Ash}
provides a detailed discussion about the relation between the bare
mass of the black hole (without hair), the mass of the hair, and the
ADM mass of the black-hole spacetime.

An increase $\Delta {\cal E}$ in the mass contained within the black
hole (due to hair swallowing) would increase the black-hole radius
from $r_H$ to $r_H+2\Delta {\cal E}$. Taking cognizance of the
black-hole area-entropy relation $S_{BH}=A/4\hbar=\pi r^2_H/\hbar$
\cite{BekTod}, one finds that the corresponding change in black-hole
entropy is given by $\Delta S_{BH}=\pi [(r_H +2\Delta{\cal
E})^2-r^2_H]/\hbar$. Substituting

\begin{equation}\label{Eq2}
\Delta S_{BH} =4\pi \Delta{\cal E} (r_H +\Delta{\cal E}) / \hbar \
,
\end{equation}
into Eq. (\ref{Eq1}), we obtain a universal lower bound on the
lifetime $\tau_{hair}$ (as measured by a static observer at
infinity) of a perturbed, unstable hairy black hole:

\begin{equation}\label{Eq3}
\tau_{hair} \geq 4(r_H +\Delta{\cal E}) \geq 4r_H \  .
\end{equation}

It is worth nothing that N\'u\~nez et. al. \cite{Nun} have proved a
very nice theorem according to which the "hairosphere", the region
where the nonlinear behavior of the black-hole hair is present, must
extend beyond ${3 \over 2}r_H$. This would imply that the collapse
time of perturbed, unstable hair into the black hole is roughly
bounded by $\geq {1 \over 2}r_H$. The universal bound, Eq.
(\ref{Eq3}), is (at least) an order of magnitude stronger. This
dynamical bound, $\tau_{hair} \geq 4(r_H +\Delta{\cal E})\geq 4r_H$,
is in fact complimentary to the spatial bound $r_{hair} \geq {3
\over 2}r_H$ of \cite{Nun}.

We would like to emphasize here that hairy black-hole solutions are
characterized by several different length/time scales:
\begin{itemize}
\item{The radius of the black-hole horizon, $r_H$.}
\item{The mass of the outside hair, ${\cal M}_{hair}$ \cite{notedimen}.}
\item{The inverse of the black-hole temperature $T_{BH}$ (proportional
to the surface gravity).}
\item{In some hairy solutions there is additional length scale set by the dimensional coupling
constant of the theory (for example, the reciprocal of the YM
coupling constant, $g$) \cite{Lavr}. Likewise, in theories with
massive hairy fields \cite{Green} there is another characteristic
scale-- the Compton length of the field, $\hbar/$mass.}
\end{itemize}

The diversity of these independent length/time scales which
characterize hairy black-hole solutions, implies that any attempt to
obtain a bound on the lifetime of hairy black holes from naive
dimensionality considerations would be too presumptuous-- there are
simply too many independent scales and dimensional parameters in the
problem. In principle, the characteristic dynamical timescale could
have turned out to be any complicated combination of the above
mentioned parameters. However, the analytically derived bound,
$\tau_{hair} \geq 4r_H$ \cite{notestro}, is remarkably simple and
universal in the sense that it is independent of all other
dimensional parameters present in the theory.

{\it Testing the bound.---} The early stages of a full, nonlinear
collapse of unstable hair can be described by linearized
perturbations. Thus the total lifetime of the unstable hairy black
hole is bounded from below by the characteristic time which
describes the early growth of linear perturbations. (The total
lifetime is actually longer and includes the phase of nonlinear
dynamics which follows the linear regime.) The linear instability
time is given by the reciprocal of the eigenvalue $\sigma$
corresponding to the unstable mode of the hairy black hole (a
linearized perturbation mode which grows according to $\sim
e^{\sigma t}$). One therefore concludes that the total lifetime of
an unstable hairy black is bounded by $\tau_{hair} \geq 1/\sigma$.
Thus, a sufficient (but not a necessary) condition for the validity
of the universal bound, Eq. (\ref{Eq3}), is that the instability
eigenvalue $\sigma$ is bounded by

\begin{equation}\label{Eq4}
\sigma \leq \sigma_{max}\equiv [4(r_H +\Delta {\cal E})]^{-1} \leq
(4r_H)^{-1}\  .
\end{equation}

As an example, we display in Table \ref{Table1} the numerically
computed (see Ref. \cite{BiCh}) eigenvalues $\sigma$ corresponding
to the unstable mode of the canonical EYM (colored) hairy black
holes. We also present the ratio $\sigma /\sigma_{max}$. We first
consider the weak version of the bound, in which case we take
$\Delta {\cal E} \to 0$. One finds $\sigma /\sigma_{max}^{weak} < 1$
for all black holes, in accord with our analytical prediction.

It has been shown \cite{Chop1,Chop2} that, depending on the initial
perturbation most of the unstable hair may collapse into the black
hole. We should therefore check the validity of our bound in its
strong version, in which case we take $\Delta {\cal E}={\cal
M}_{hair}$. We display the ratio $\sigma /\sigma_{max}^{strong}$ in
Table \ref{Table1}, from which one learns that the unstable hairy
black holes conform to the bound Eq. (\ref{Eq4}) even in its strong
form. (In fact, the colored black holes are actually very close of
saturating this dynamical bound.) It is worth mentioning that the
EYM solutions have a second type of instability (``sphaleron" or
topological one), which may develop outside the magnetic ansatz to
which the solution itself belongs, see \cite{Lav1,Lav2}. These two
types of instabilities have comparable spectrums and are both
relevant for the lifetimes of the EYM hairy black holes.

\begin{table}[htbp]
\centering
\begin{tabular}{|c|c|c|c|c|}
\hline
$r_H$ & ${\cal M}_{hair}$ & $\sigma$ & $\sigma /\sigma_{max}^{weak}$ & $\sigma /\sigma_{max}^{strong}$ \\
\hline
\ 0.0\ \ & \ 0.8286\ \ &\ 0.2292\ \ & 0.000 & 0.759 \\
\ 0.4\ \ & \ 0.6589\ \ &\ 0.2071\ \ & 0.331 & 0.877 \\
\ 0.6\ \ & \ 0.5791\ \ &\ 0.1936\ \ & 0.465 & 0.913 \\
\ 1.0\ \ & \ 0.4369\ \ &\ 0.1637\ \ & 0.655 & 0.941 \\
\ 2.0\ \ & \ 0.2349\ \ &\ 0.1067\ \ & 0.854 & 0.954 \\
\ 3.0\ \ & \ 0.1571\ \ &\ 0.0744\ \ & 0.893 & 0.940 \\
\ 4.0\ \ & \ 0.1168\ \ &\ 0.0562\ \ & 0.899 & 0.925 \\
\ 5.0\ \ & \ 0.0923\ \ &\ 0.0447\ \ & 0.894 & 0.911 \\
\hline
\end{tabular}
\caption{Ratio between the eigenvalue $\sigma$ corresponding to the
unstable mode of the colored hairy black holes and $\sigma_{max}$,
as defined by the bound Eq. (\ref{Eq4}). Here $\sigma_{max}^{weak}
\equiv (4r_H)^{-1}$ and $\sigma_{max}^{strong} \equiv [4(r_H+ {\cal
M})_{hair}]^{-1}$. One finds $\sigma/\sigma_{max} < 1$ for all
values of the black-hole radius $r_H$, in accord with the analytical
prediction, Eq. (\ref{Eq4}).} \label{Table1}
\end{table}

{\it Discussion and applications.---} The universal laws of
thermodynamics and information theory provide important insights on
the dynamics of unstable hairy black holes. In particular, we have
derived a simple lower bound on the lifetime of perturbed, unstable
hair: {\it If a black hole has hair, then its lifetime cannot be
shorter than $4$ times the horizon radius}. This temporal bound,
$\tau_{hair} \geq 4(r_H+{\cal M}_{hair}) \geq 4r_H$ is complimentary
to the spatial bound $r_{hair} \geq {3 \over 2}r_H$ of \cite{Nun}.
It would be highly interesting to verify directly the validity of
the analytic bound, Eq. (\ref{Eq3}), with full nonlinear dynamical
calculations.

It is worth pointing out that, although our starting point, Eq.
(\ref{Eq1}) has clear quantum origins, the analytically derived
bound on the dynamical lifetime of hairy black holes, Eq.
(\ref{Eq3}), is of pure classical nature. [The $\hbar$ that appears
in (\ref{Eq1}) is canceled out by the $\hbar$ in the black-hole
area-entropy relation, $S_{BH}=A/4\hbar$.] This strongly suggests
that one should be able to obtain a bound on the lifetime of an
unstable black hole by using purely classical arguments. One such
(analytic) example is given in Appendix A, where we consider the
specific case of the SU(2) Reissner-Nordstr\"om black hole. However,
such approach would be parameter and model dependent. One would be
forced to go into details about the system, and to solve the problem
case-by-case for each and every theory in which black-hole hair has
been discovered. On the other hand, the inequality (\ref{Eq1}) is
remarkably robust-- it therefore enables a simple derivation of a
universal, model independent bound like (\ref{Eq3}).

If the central black hole is small, then the dynamics of its outside
fields would hardly be affected by its presence. In such cases, the
dimensional parameters of the matter fields (e.g., the dimensional
coupling constant, $g$, of the YM theory) would determine the
characteristic dynamical timescale of the outside hair. It is
interesting to note that if we take this natural expectation and
combine it with the temporal bound $\tau \geq 4(r_H +{\cal
M}_{hair})$, we obtain an upper bound for the mass of the black-hole
hair outside the horizon. Namely, substituting $\tau \sim g^{-1}$
for the characteristic dynamical timescale would yield,

\begin{equation}\label{Eq5}
{\cal M}_{hair} \lesssim 1/g-r_H\ \ \ ;\ \ \ r_H \lesssim 1/g\ .
\end{equation}
This bound implies that a black hole cannot support an unbounded
amount of hair. The mass of the outside hair is fundamentally
bounded by the reciprocal of the coupling constant of the theory.
Moreover, the amount of allowed hair decreases with increasing
black-hole radius. These analytical conclusions are in accord with
available numerical data, see e.g., \cite{BiCh}. The transition from
hairy black-hole solutions which are dominated by their outside hair
(${\cal M}_{hair} \gg r_H$), to solutions dominated by the central
black hole (${\cal M}_{hair} \ll r_H$) occurs at $r_H \sim 1/g$.

Likewise, in theories with massive hairy fields there is a natural
length/time scale in the problem set by the Compton length of the
field, $\hbar/m$, where $m$ is the mass parameter. This scale would
characterize the dynamics of the outside fields if the central black
hole is small. Taking $\tau \sim \hbar/m$ in the relation $\tau \geq
4(r_H +{\cal M}_{hair})$, one obtains an upper bound on the mass of
the outside hair,

\begin{equation}\label{Eq6}
{\cal M}_{hair} \lesssim \hbar/m -r_H\ \ \ ;\ \ \ r_H \lesssim
\hbar/m\ .
\end{equation}
This analytical bound is in accord with known numerical data
\cite{Nun,Green}, and implies that ${\cal M}_{hair} \to 0$ as $r_H
\simeq \hbar/m$.

The above upper bounds do not rule out the existence of hairy
black-hole solutions. However, they put a severe restriction on the
amount of outside hair that a black hole can support. Such a
restrictive upper bound can therefore be taken as a more modest
alternative to the original no hair conjecture.

Finally, we would like to point out that our approach sheds some new
light on the black-hole critical phenomena observed in full
nonlinear gravitational collapse \cite{ChopCr}. It has been found
\cite{Chop1,Chop2} that unstable hairy black holes may serve as
intermediate attractors during a dynamical gravitational collapse.
Near-critical evolutions approach the solution of a static hairy
black hole, and remain in its vicinity for some amount of time $T$.
The evolution then peels off the static hairy black hole solution,
with the remaining field (outside of the horizon) either dispersing
to infinity or collapsing and adding a finite amount of mass to the
already present black hole \cite{Chop1,Chop2}. The lingering time,
$T$, in the vicinity of the intermediate attractor (the unstable
hairy black-hole solution) is related directly to the reciprocal of
the instability eigenvalue $\sigma$ \cite{BiCh}. Our results
therefore set an analytic lower bound on the lifetime, $T$, of
near-critical solutions in Choptuik's nonlinear critical phenomena.

\newpage
\setcounter{equation}{0}
\renewcommand{\theequation}{A\arabic{equation}}
\begin{appendix}\label{App1}
{\bf Appendix A: Instability of the SU(2) Reissner-Nordstr\"om black
hole}

In this appendix we consider the linearized stability analysis of
the SU(2) Reissner-Nordstr\"om (RN) black hole with unit magnetic
charge \cite{Yass}. (This black-hole solution is not regarded as
hairy since its outside mass is related to the global magnetic
charge. Nevertheless, we present here the stability analysis of this
black hole because it has the advantage of being tractable
analytically.)

The mass function of the SU(2) RN black hole of asymptotic mass $M$
is given by

\begin{equation}\label{EqA1}
m(r)=M-{1 \over {2r}}\  .
\end{equation}
This implies that the black-hole horizon is located at
$r_H=M+(M^2-1)^{1/2}$.

The evolution of linearized perturbations $\xi(r) e^{\sigma t}$ is
governed by a Schr\"odinger-like equation \cite{Bizw}

\begin{equation}\label{EqA2}
\Big( -{{d^2} \over {dx^2}} +U(x) \Big)\xi=-\sigma^2 \xi \  ,
\end{equation}
where $dx/dr=[1-2m(r)/r]^{-1}$ (this implies $x \to -\infty$ for $r
\to r_H$, and $x \to \infty$ for $r \to \infty$), and the effective
potential is given by

\begin{equation}\label{EqA3}
U(x)=-{1 \over {r^2}}\Big(1-{{2m(r)} \over r}\Big) \  .
\end{equation}
Eq. (\ref{EqA2}) has at least one bound state, because the effective
potential $U(x)$ is everywhere negative and vanishes for $|x| \to
\infty$. Moreover, an upper bound on $\sigma^2$ is given by the
(minus) minimum of the potential $U$ \cite{Bizw}. Namely, $\sigma
\leq \sqrt{|U_{min}(M)|}$. From here one finds that $\sigma$ is
bounded according to

\begin{equation}\label{EqA4}
\sigma(M) < [4\beta(M)r_H]^{-1}\  ,
\end{equation}
where $\beta(M)$ is a monotonic decreasing function from
$\beta(M=1)=1$ to $\beta(M \to \infty)=\sqrt{27}/8$. This yields a
lower bound on the black-hole lifetime, $\tau \geq 1/\sigma$ which
is a bit weaker than our universal bound, Eq. (\ref{Eq3}).

\end{appendix}
\maketitle

\bigskip
\noindent {\bf ACKNOWLEDGMENTS}
\bigskip

This research is supported by the Meltzer Science Foundation. I
thank Piotr Bizo\'n for helpful correspondence. I also thank Uri
Keshet and Liran Shimshi for stimulating discussions.


\begin{thebibliography}{99}
\bibitem{wec} We assume the validity of the weak energy condition,
which insures the positivity of the hair's mass.
\bibitem{notedimen} We emphasize that the mass of the black-hole hair, ${\cal M}_{hair}$ does not
scale linearly with its raduis, $r_H$, see e.g. \cite{BiCh}. These
two quantities therefore determine different length/time scales.
\bibitem{notestro} The bound can be made stronger, $\tau_{hair} \geq
4(r_H+{\cal M}_{hair})$, as discussed below.
\end{thebibliography}
\end{document}